\documentclass[11pt]{article}
	\usepackage{amsmath}
	\usepackage{amssymb}
	\usepackage{graphicx}
	\usepackage{hyperref}

	\def\aa{A\&A }
	
	\def\aj{AJ }
	\def\apj{ApJ}
	\def\apjs{ApJS}
	\def\apjl{ApJL}

	\def\mnras{MNRAS }

\begin{document}
	
	\setcounter{figure}{0}
	\setcounter{table}{0}
	\setcounter{footnote}{0}
	\setcounter{equation}{0}
	
	\noindent {\Large\bf THE FUNDAMENTAL REFERENCE AGN MONITORING EXPERIMENT (FRAMEX)}
	\vspace*{0.7cm}

	\noindent\hspace*{1cm} B. DORLAND$^1$, N. SECREST$^2$,  M. JOHNSON$^3$, T. FISCHER$^4$, N. ZACHARIAS$^5$, J. SOUCHAY$^6$, S. LAMBERT$^7$, C. BARACHE$^8$, F. TARIS$^9$ \\[0.2cm]
	\noindent\hspace*{1cm} $^1$ U.S. Naval Observatory, Washington, DC - USA - bryan.dorland@navy.mil\\
	\noindent\hspace*{1cm} $^2$ U.S. Naval Observatory, Washington, DC - USA - nathan.secrest@navy.mil\\
	\noindent\hspace*{1cm} $^3$ U.S. Naval Observatory, Washington, DC - USA - megan.johnson@navy.mil\\
	\noindent\hspace*{1cm} $^4$ Space Telescope Science Institute, Baltimore, MD - USA - tfischer@stsci.edu\\
	\noindent\hspace*{1cm} $^5$ U.S. Naval Observatory, Washington, DC - USA - norbert.zacharias@navy.mil\\
	\noindent\hspace*{1cm} $^6$ Observatoire de Paris, PSL Research
University, UMR8630 CNRS - France - \linebreak jean.souchay@obspm.fr\\
	\noindent\hspace*{1cm} $^7$ Observatoire de Paris, PSL Research
University, UMR8630 CNRS - France - \linebreak sebastien.lambert@obspm.fr\\
	\noindent\hspace*{1cm} $^8$ Observatoire de Paris, PSL Research
University, UMR8630 CNRS - France - \linebreak christophe.barache@obspm.fr\\
	\noindent\hspace*{1cm} $^9$ Observatoire de Paris, PSL Research
University, UMR8630 CNRS - France - \linebreak francois.taris@obspm.fr\\

	\vspace*{1cm}
	
	\noindent {\large\bf ABSTRACT.} The U.S. Naval Observatory (USNO), in collaboration with Paris Observatory (OP), is conducting the Fundamental Reference AGN Monitoring Experiment, or FRAMEx.  FRAMEx will use USNO's and OP's in-house observing assets in the radio, infrared (IR) and visible, as well as other ground- and space-based telescopes (e.g., in the X-ray) that we can access for these purposes, to observe and monitor current and candidate Reference Frame Objects (RFOs)---consisting of Active Galactic Nuclei (AGN)---as well as representative AGN, in order to better understand astrometric and photometric variability at multiple timescales.  FRAMEx will improve the selection of RFOs as well as provide significant new data to the AGN research community.  This paper describes the FRAMEx objectives, specific areas of investigation, and the initial data collection campaigns.  
	
	\vspace*{1cm}
	
	\noindent {\large\bf 1. INTRODUCTION AND BACKGROUND}
	\smallskip
	
	The International Celestial Reference System (ICRS) is the standard reference system for defining position and motion of celestial objects.  It is currently realized by the International Celestial Reference Frame 3 (ICRF3), which consists of Very Long Baseline Interferometry (VLBI) measurements of a total of 4536 extragalactic radio sources (Charlot, et al. 2020).  These are primarily Active Galactic Nuclei (AGN), which are Supermassive Black Holes (SMBH) at the centers of galaxies, with a typical redshift of $z=1.0$.  (Fey \& Charlot 2000, Charlot, et al. 2020, Souchay, et al. 2019)
	
ICRF3 was adopted as the international standard by the International Astronomical Union (IAU) on 1 January 2019 (Lago 2019).  It is the first multi-wavelength reference frame to be adopted by the IAU, with observations in S/X (4536 sources), K (824 sources), and X/Ka (674 sources).  Of these sources, 303 are considered ``defining sources,'' which are distributed quasi-isotropically over the entire sky, and set the rotation of the reference frame.  The remaining sources are used to densify the resultant reference frame.  ICRF3 is assessed to have a mean defining source position uncertainty of 30 $\mu$as in DEC and 3 $\mu$as in RA, with a per-source positional noise floor of 30 $\mu$as (Charlot, et al. 2020).

Nearly simultaneous with the release of ICRF3, the European Space Agency (ESA) Gaia mission, using AGN previously identified in the IR from NASA's Wide-field Infrared Survey Explorer (WISE) mission (Secrest, et al. 2015), produced an optical catalog of over 500k AGN (including ICRF3 sources) accurate to the 1 mas level or better (Mignard, et al. 2018).  Ultimately, Gaia aims to produce an optical reference frame at the few tens of $\mu$as accuracy per source, comparable to the ICRF3.  

With realizations of the ICRS now effectively in four different wavelengths, it will be the responsibility over the next decade, as new data are taken,  of the Celestial Reference Frame community---including the IAU and the International Earth Rotation and Reference Systems Service (IERS)---to ensure that the integrated reference frame is both accurate and aligned properly across wavelengths.  As new and improved instrumentation becomes available for observing reference frame objects, these new capabilities should be deployed to identify the most stable and accurate candidate objects across wavelengths, and to deselect objects that are unstable or otherwise problematic.  In doing so, a much better understanding of the underlying astrophysics of AGN will be enabled.  


	
	
	\vspace*{0.7cm}
	\noindent {\large\bf 2. REFERENCE FRAME OBJECTS: ISSUES AND QUESTIONS}
	\smallskip
	
	As described in section 1, the underlying reference frame is realized by over four thousand reference frame objects (RFOs) (i.e., AGN) in the radio observed at three different frequencies, and half a million observed in the visible.  How accurate are these observations?  How do positions measured in one wavelength compare to positions measured in a different one?  These positions were measured in different epochs using different classes of instruments: how do the positions change in time due to physical changes of the sources, differences at the measurement instrument, or differences in observing epochs?  
	
	A significant amount of work has already gone into exploring issues associated with apparent offsets of astrometric positions measured in different wavelengths. These include Zacharias \& Zacharias (2015), which observed a number of positional offsets between radio and optical positions that significantly exceeded ($>3 \sigma$) offsets expected due to measurement errors; Makarov et al. (2019) found that 20\%, rather than the expected 1\%, of RFOs exceeded a normalized radio-optical offset of 3; Roland, et al. (2019) argued for the effects of binary black holes on the apparent radio position ``noise'' of RFOs; and Petrov, et al. (2019) found a significant correlation between radio-optical offsets and the direction of the ostensible AGN jet as observed in the radio.  These all point to aspects of the problem of variability in position between wavelengths and in time, and to possible ``clutter'' causing offsets in the measured positions of AGN in one or more wavelengths.  
	
	The goal of the \underline{F}undamental \underline{R}eference Frame \underline{A}GN \underline{M}onitoring \underline{Ex}periment (FRAMEx)  is to observe and constrain the astrometric position of RFOs by monitoring both the astrometry and photometry of RFOs and other, representative AGN across multiple wavelengths, spanning extended time periods, and at a variety of different temporal frequencies.  FRAMEx will probe possible correlations between photometric variability across the spectrum and astrometric position variability.  Specific areas of issues and questions to be probed include:
	
	\begin{itemize}
  	\item \textbf{Source of RFO/AGN emissions.}  What physical processes are responsible for the emissions observed at different wavelengths?  What are the magnitudes of the 
	offsets measured in different bands or frequencies?  What are the dependencies of these positions and offset considerations such as the resolving power or sensitivity of the 
	instrument?  What are contributions from possible sources of offset such as:
	
	\begin{itemize}
		\item Confused foreground or background sources of emission and \emph{in situ} sources of offset such as off-nuclear AGN, host galaxy structure and brightness
		\item Confusion or blending of discrete sources
		\item Different sources of emission as a function of wavelength
		\item Narrow Line Region (NLR) emission and AGN outflows
		\item Wavelength-dependent line-of-sight AGN obscuration
		\item Emission from extended radio jets.	
		
		\end{itemize}
	
	\item \textbf{Photometric and astrometric variability.}  What is the time-dependent astrometric and photometric behavior of the sources of AGN emission defined previously?  What 
	timescales are relevant for variability?  What are the correlations between astrometric and photometric variability, and what does this tell us about what is going on at the AGN, in 
	the host galaxy environment, the intervening intergalactic medium, or even our local observing conditions? 
		
	\item \textbf{Binary/Multiple SMBH population, orbital characteristics, and resultant phenomenology.}  What is the population of binary (or multiple) SMBHs?  For binary or multiple 
	SMBHs, what critical phenomenologies reveal their presence or allow us to determine their physical parameters?  What are the range of parameters (e.g., orbital periods and 
	distances) observed?  Are there binary systems in which the VLBI counterpart is one AGN while the Gaia counterpart is the other?	
	
	\item \textbf{Possible RFO proper motion.} What is the explanation of the (apparent) large proper motion of AGN as observed in Gaia DR2?
	\end{itemize}
	
	\vspace*{0.7cm}
	\noindent {\large\bf 3. FRAMEX AGN AND HOST GALAXY DISTANCE SCALES}
	\smallskip
	
	In order to understand what can and can't be observed (and, therefore, what astrophysics can be probed by FRAMEx), we first consider the spatial scales associated with the observations.  As noted in section 1, the ``typical'' AGN used as an RFO is at $z=1.0$.  This translates to an angular size distance $D_A \approx 5\times10^{9}$ light years (see various cosmological calculators, such as \url{http://www.astro.ucla.edu/~wright/CosmoCalc.html} and associated references).  
	
	Typical distances for a ``Milky Way''-type galaxy are shown in Fig. 1, along with the angles subtended for an angular size distance $D_A \approx 5\times10^{9}$ light years corresponding to $z=1.0$.  The inset in the upper center shows the neighborhood around the AGN with the major features of the AGN noted.  Distances for the features are taken from Ricci (2020).  To put this configuration in scale, the event horizon is solar-system sized; the accretion disk is the size of the Sun's Oort cloud, and the torus is on the scale of stellar distances in our local region of the Galaxy.  The Narrow Line Region is much more extended, going out to as far as $\approx300$ pc from the central AGN.  Different radiative processes (both emission and absorption) are associated with each of these features, and all of them will contribute or affect the measured position of the AGN as a function of both wavelength and time.
	
	Beyond the AGN features, the scale of the host galaxy is shown.  The central bulge is about 2 kpc, the disk diameter is 20 kpc, and the halo is 30 kpc.  Satellite galaxies are approximately 50 kpc from the core of the host galaxy of the AGN, and nearby galaxies of comparable size are 800 kpc distant.  Each one of these regions may contain sources of emission or absorption that affect the measured position of the central AGN as a function of both wavelength and time.  
	
	Table 1 captures this information for each of the AGN and host galaxy features.  Because these features span nine orders of magnitude in distance, different ``natural units'' are associated with the different features.  These are included in the table where they are considered useful.  Also included is the subtended angle, calculated using $D_A$.  The final two columns on the right indicate what instruments, if any, are capable of measuring either the astrometric position of the AGN or resolving the relevant feature.  Photometric accuracy is driven primarily by the sensitivity of the instrument and integration time of the observation, but is also affected by blending of sources that only high resolution may be able to resolve.
	
	Figure 2 displays the orbital periods for three binary AGN configurations as a function of apparent angular separation for $z=1.0$.  The three configurations consist of $10^6$, $10^9$ and $10^{10} M_\odot$ pairs of SMBHs, respectively.  The relevant features from Fig. 1 and Table 1 are shown along the top, and representative instrument capabilities for astrometric measurements and resolved images.  This figure clarifies a few key concepts: first, orbital motion will only be detectable over the timescales accessible to FRAMEx for the largest and closest pairs of binary AGN.  Second, the array of instruments available to us for probing scales down to the NLR is limited; spatial scales for the AGN (e.g., torus, accretion disk, event horizon), will require new, more precise instrumentation to be developed.  VLBI in general and the VLBA specifically are critical to probing spatial scales smaller than the bulge, and are able to resolve components that would be blended in other types of observations.  VLBI, because of the much narrower resolution element, also offers the highest astrometric resolution for single measurements, being sensitive to measurements at scales that probe the narrow line region. 
	
	\begin{figure}[h]
	\begin{center}
	\centerline{
	\includegraphics[scale=0.7]{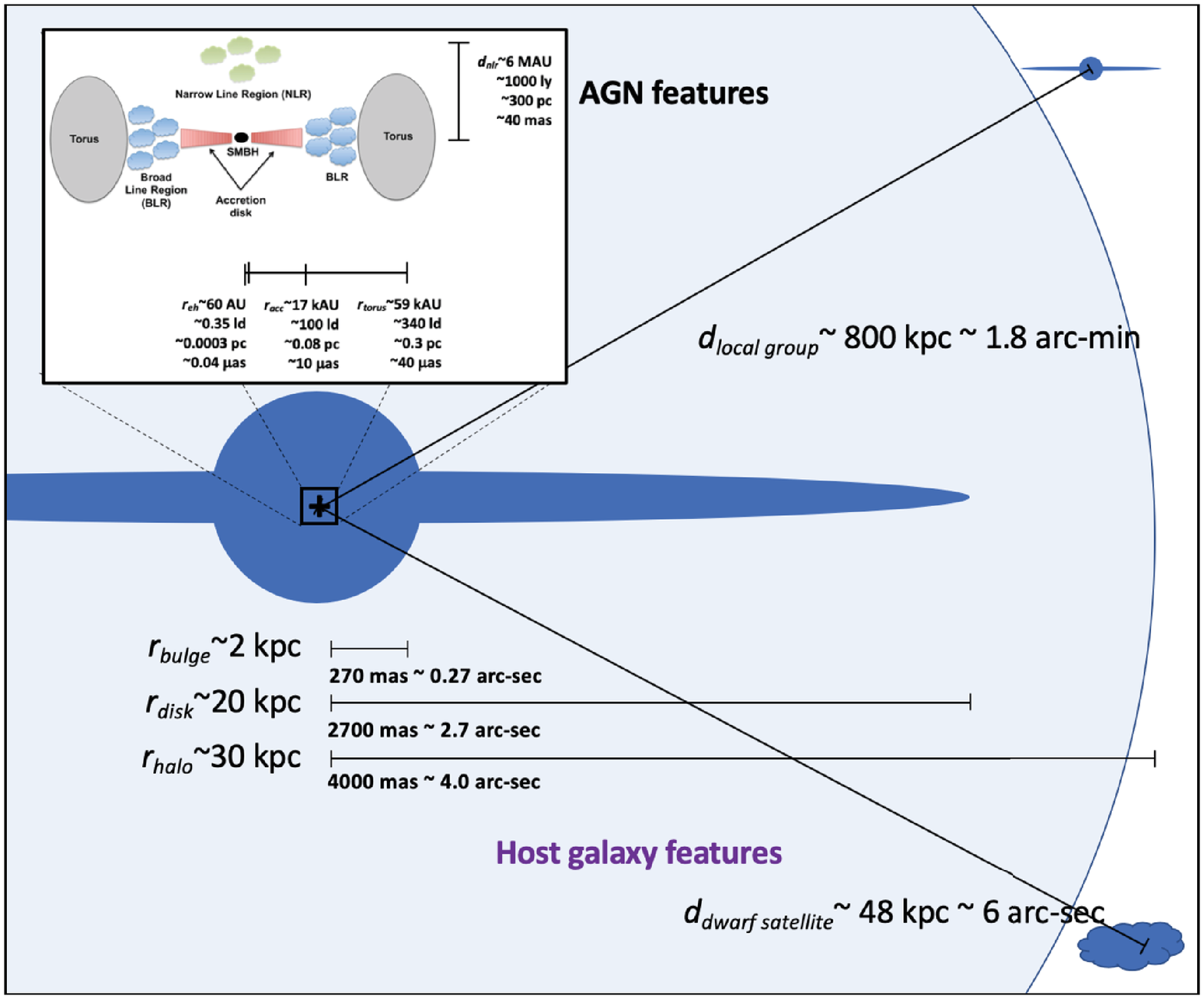}
	}
	\caption{Relevant distances scales for ``typical'' ($z=1.0$) RFO AGN.  Angular equivalents assume a distance of $5\times10^{9}$ light years. Host galaxy similar to Milky Way is shown, with central box indicating the AGN core region.  Figure not to scale. Inset taken from C. Ricci's website, \url{http://www.isdc.unige.ch/~ricci/Website/Active_Galactic_Nuclei.html}}
	\end{center}
	\end{figure}

	\begin{table}[h]
	\begin{center}
	 \begin{tabular}{l | c | c | c | c | c | c | c} 
      		\textbf{Feature} & \multicolumn{5}{c |}{\textbf{Distance}} &\multicolumn{2}{c}{\textbf{Instrumentation}}\\ 
		\hline
      		 (parameter) & AU & light & light & parsecs & angle & to measure & to  \\ 
		                     &       & days & years &   & subtend. & position & resolve \\
      		\hline\hline
      			Event hor. (r) 	& $60$ 			& 0.35 	&  				&  $3\times 10^{-4}$ 		& $0.04 \mu$as &  				& \\
      			Acc. disk (r)	& $1.7 \times 10^4$ 	& 100 	& 0.3 			& $8\times 10^{-2}$ 		& $10 \mu$as  	&  				& (*) \\
     			Torus (r) 		& $5.9 \times 10^4$ 	& 300 	& 0.8 			& $2.4\times 10^{-1}$ 	& $40 \mu$as 	& 1			& \\
			NLR (s)		& 			 	&  		& $10^3$ 			& $3\times 10^{2}$ 		& 40mas 		& 1, 2, 3, 4 	& 1\\
    			Bulge (r) 		&  				&  		& $6.5\times 10^3$ 	& $2\times 10^{3}$ 		& 270mas		& 1, 2, 3, 4 	& 1, 2, 3, 4$_{AO}$ \\
      			Disk (r)		&  				&  		& $6.5\times 10^4$ 	& $2\times 10^{4}$ 		& 2.7'' 		& 1, 2, 3, 4	& 2, 3, 4\\
     			Halo (r)		&  				&  		& $9.8\times 10^4$	& $3\times 10^{4}$ 		& 4'' 			& 1, 2, 3, 4	& 2, 3, 4\\
			Sat. dwarf (s)&  				&  		& $1.6\times 10^5$ 	& $4.8\times 10^{4}$ 	& 6'' 			& 1, 2, 3, 4	& 2, 3, 4\\
			Local grp. (s) &  				&  		& $2.6\times 10^6$ 	& $8\times 10^{5}$ 		& 1.8' 			& 1, 2, 3, 4	& 3, 4\\
   	 \end{tabular}
	\caption{Typical distances for AGN-related and host galaxy features for a Milky Way-type galaxy in a ``Local Group''-like cluster, from Fig.1.  Features are show in column 1, with (r) indicating a radius, and (s) a separation. Distances are shown in columns 2--6, with the equivalent angular plane-of-sky separation shown in column 6.  Columns 3 and 4 refer to ``light days'' and ``light years''.  Columns 7 and 8 indicate the class of astronomical telescope/array needed to make the observation.  Column 7 is for single-epoch astrometric position measurement, and column 8 is to resolve the RFO feature.  1 indicates VLBA, 2 is VLA, 3 is space-based optical, and 4 is ground-based optical.  A subscript ``AO'' indicates where adaptive optics is required to reach the indicated resolution from the ground.  The asterisk indicates the specialized ``Event Horizon Telescope'' (EVT), a world-wide Very Long Baseline Interferometric demonstration array that was able to image the accretion disk of the M87 SMBH in 2019 (Event Horizon Telescope Collaboration, et al., 2019).}
	\end{center}
	\end{table}

	\begin{figure}[h]
	\begin{center}
	\includegraphics[scale=0.53]{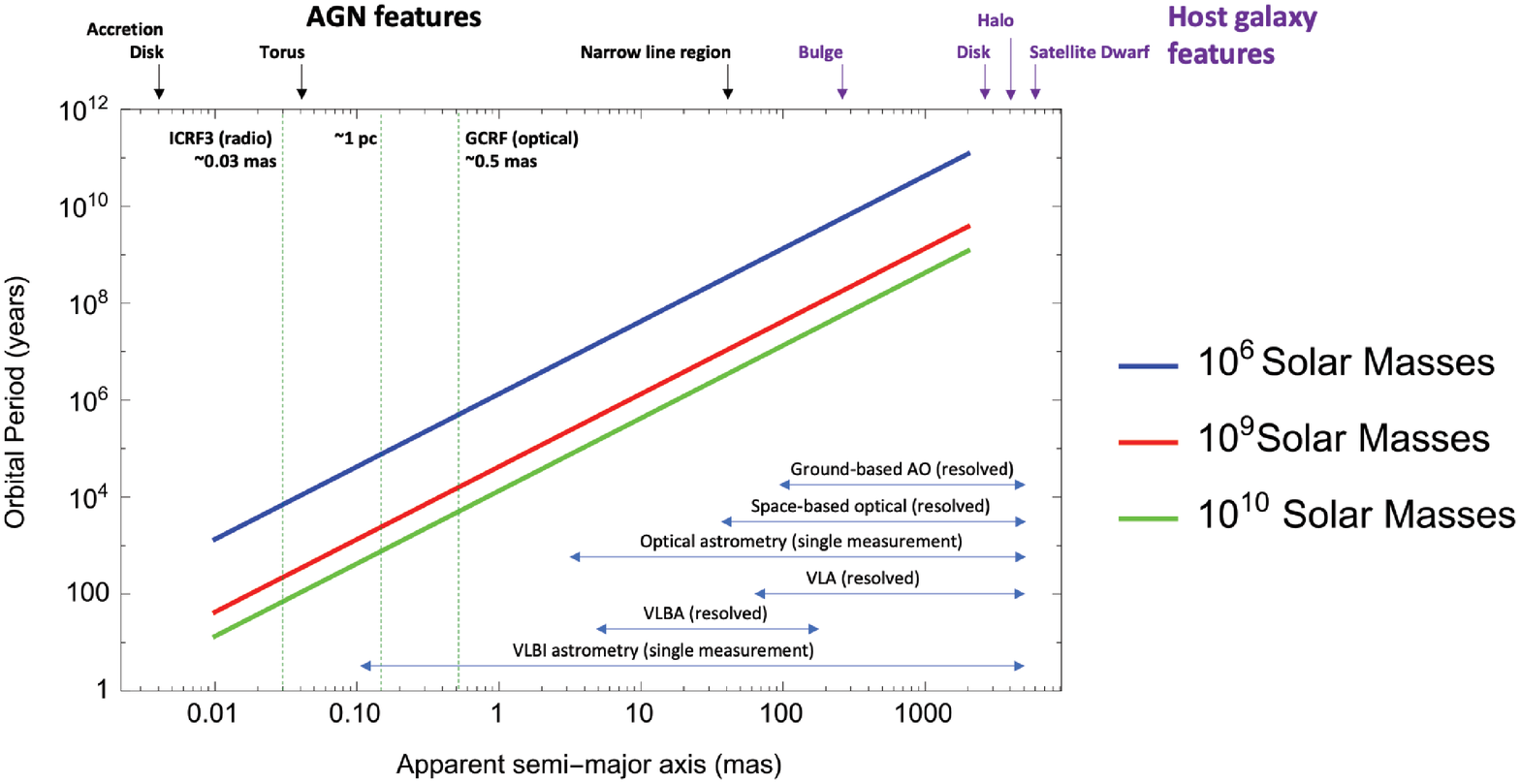}
	\caption{Apparent angular separation vs. orbital period, shown for three binary SMBH configurations: $10^{6}$, $10^9$, and $10^{10}$ solar mass components. Relevant AGN (black) and Host galaxy features (purple) (see Fig.\ 1 and Table 1) are shown along the top of the plot.  Also shown, in lower right, are approximate equivalent observing capabilities of  radio and visible telescopes and arrays.  ICRF3 and Gaia Celestial Reference Frame (GCRF) accuracies are shown as dashed green lines.  Also show is the equivalent 1 pc plane-of-the-sky distance at $z=1.0$.}
	\end{center}
	\end{figure}

	\vspace*{0.7cm}
	\noindent {\large\bf 4. INITIAL AREAS OF INVESTIGATION}
	\smallskip
	
	FRAMEx will utilize the instrumentation available to the U.S. Naval Observatory and the Paris Observatory, as part of the collaboration activity.  This includes 1 and 2-m class optical and near infrared (NIR) telescopes operated by both USNO and OP in a variety of locations, as well as both the UK Infrared Telescope (UKIRT) and the Very Long Baseline Array (VLBA).  Additional observations will be conducted using instruments that are accessed on a case-by-case basis, including both ground- and space-based telescopes requiring proposals.
	
	The initial areas of investigation will focus on the following:
	
	\begin{itemize}
  	\item \textbf{Volume-limited sample of AGN}  The ``volume limited AGN sample'' will concentrate on $\approx 25$ of the nearest AGN that are visible from the Northern Hemisphere.  Observations will characterize these AGN at spatial resolutions that are not available for more distant AGN, on the assumption that the results will be extensible.  The volume-limited approach significantly reduces biases that result from selection effects.  The goal will be to observe this limited sample in visible, IR, radio and x-ray.  These multi-wavelength observations that are conducted at the same epoch for this set of targets will be a unique set of data for AGN and RFO research.
	\item \textbf{Problematic ICRF Defining Sources}  We will identify approximately three dozen ``problematic'' ICRF defining sources that display statistically significant offsets between S/X and either K or X/Ka positions or with Gaia optical positions.  We will also identify a comparable number of non-problematic sources.  We will monitor both of these groups in the radio, visible and IR for astrometric or photometric variability, and use these observations to identify possible astrophysical explanations of the offsets.
	\end{itemize}
	
	In addition to these areas of investigation, we plan to complement FRAMEx with the following activities:
	
		\begin{itemize}
  	\item \textbf{Redshift survey for all RFOs and candidate RFOs.}  Approximately half of RFOs have traceable spectroscopic redshifts, mostly from the Sloan Digital Sky Survey.   Observing priority will be on ICRF3 defining sources, but the goal is to survey all ICRF3 RFOs for which we do not currently have a measured redshift.  
	\item \textbf{ICRF Source Structure Imaging Database.}  Associated with FRAMEx, and in support of USNO agreements with the International VLBI Service for Geodesy and Astrometry (IVS) and IERS agreements, we are in the process of developing a data archive of VLBA and VLBI images of AGN (the Fundamental Reference Image Data  Archive, or FRIDA (Hunt 2019, 2020).  These high-resolution images will be made available to the public, and will eventually be supplemented by other data products (including any visible, IR, or x-ray observations) associated with the relevant RFOs.
	\item \textbf{Southern Hemisphere 1-m Monitoring Survey.}  USNO's 1-m visible and NIR Deep South Telescope (DST) is being dedicated primarily to observing RFOs.  Target lists concentrate on astrometric and photometric variability across this spectral regime for RFOs.  These data will not only support FRAMEx work, but be used to populate the FRIDA to support general AGN and RFO research and development activities.  
	\item \textbf{Multi-wavelength, highly stable RFO selection}  One by-product of the FRAMEx work to identify AGN variability will be an identification of those AGN that are either not, or minimally, variable across wavelengths (assuming there are such AGN).  These AGN can then be used as ``pre-screened candidates'', with well-characterized offsets and well-understood stability parameters for use in the development of future reference frames. 
	\end{itemize}

	\vspace*{0.7cm}
	\noindent {\large\bf 5. CONCLUSION}
	\smallskip
	
	USNO, in collaboration with Paris Observatory and other partners, is moving forward with a dedicated investigation of reference frame objects (RFOs) using both in-house and external observing assets.  The multi-year collaboration, called FRAMEx, will concentrate on ICRF3 sources, but also seek to observe other AGN and quasar sources so as to better understand the underlying astrophysics and improve the reference frame across the spectrum going forward.  	

	

	%
	%
	%
	%

	\vspace*{0.7cm}
	\noindent{\large\bf 7. REFERENCES}
	{
	
	\leftskip=5mm
	\parindent=-5mm
	\smallskip
	
	Charlot, P., et al.\ 2020, ``International Celestial Reference Frame 3'', submitted
	
	Event Horizon Telescope Collaboration, Akiyama, K., Alberdi, A., et al.\ 2019, ``First M87 Event Horizon Telescope Results. I. The Shadow of the Supermassive Black Hole'', \apjl, 875, L1
	
	Fey, A.~L., \& Charlot, P.\ 2000, ``VLBA Observations of Radio Reference Frame Sources. III. Astrometric Suitability of an Additional 225 Sources'', \apjs, 128, 17
	
	Hunt, L., Johnson, M., Fey, A.~L., Gordon, D., Spitzak,  2019, ``VLBA Imaging of ICRF3 Sources'', these proceedings, \url{https://syrte.obspm.fr/astro/journees2019/FILES/johnson_ea.pdf}
		
	Hunt, L., Fey, A.~L., Gordon, D., Spitzak, J.\ 2020, ``Imaging over 3,000 Quasars that are a part of the International Celestial Reference Frame'', American Astronomical Society Meeting Abstracts \#235
	
	Lago, T. (ed.) 2019, ``RESOLUTION B2 on The Third Realization of the International Celestial Reference Frame'', Proc. XXX IAU General Assembly,  Volume XXXB
	
	Makarov, V.~V., Berghea, C.~T., Frouard, J., et al.\ 2019, ``The Precious Set of Radio-optical Reference Frame Objects in the Light of Gaia DR2 Data'', \apj, 873, 132
	
	Mignard, F., Klioner, S.~A., et al.\ 2018, ``Gaia Data Release 2. The celestial reference frame (Gaia-CRF2)'', \aa, 616, A14
	
	Petrov, L., Kovalev, Y.~Y., \& Plavin, A.~V.\ 2019, ``A quantitative analysis of systematic differences in the positions and proper motions of Gaia DR2 with respect to VLBI'', \mnras, 482, 3023
	
	Ricci, C. 2020, ``Active Galactic Nuclei'', \url{http://www.isdc.unige.ch/~ricci/Website/Active_Galactic_Nuclei.html}
	
	Roland, J., Gattano, C., Lambert, S., \& Taris, F.\ 2019, ``Structure and variability of quasars'', these proceedings, \url{https://syrte.obspm.fr/astro/journees2019/journees_pdf/SessionIII_2/ROLAND_Jacques.pdf}
	
	Secrest, N., Dudik, R., Dorland, B.~N., et al. 2015, ``Identification of 1.4 Million Active Galactic Nuclei in the Mid-Infrared using WISE Data'', ApJ Supplement, 221, 12S
	
	Souchay, J., Gattano, C., Andrei, A., et al. 2019, ``LQAC-5: The fifth release of the Large Quasar Astrometric Catalogue'', \aa, 624, A145 
	
	Zacharias, N., \& Zacharias, M.~I.\ 2014, ``Radio-Optical Reference Frame Link Using the U.S. Naval Observatory Astrograph and Deep CCD Imaging'',  \aj, 147, 95

	}
	
	\end{document}